# Secure Email - A Usability Study

Adrian Reuter[1], Karima Boudaoud[2], Marco Winckler[2], Ahmed Abdelmaksoud[2], Wadie Lemrazzeq[2]

[1] Technische Universität München, Arcisstraße 21, 80333 München, Germany
[2] Polytech Nice Sophia, 930 route des Colles, 06903 Sophia Antipolis, France

**Abstract.** Several end-to-end encryption technologies for emails such as PGP and S/MIME exist since decades. However, end-to-end encryption is barely applied. To understand why users hesitate to secure their email communication and which usability issues they face with PGP, S/MIME as well as with pEp (Pretty Easy Privacy), a fairly new technology, we conducted an online survey and user testing. We found that more than 60% of e-mail users are unaware of the existence of such encryption technologies and never tried to use one. We observed that above all, users are overwhelmed with the management of public keys and struggle with the setup of encryption technology in their mail software. Even though users struggle to put email encryption into practice, we experienced roughly the same number of users being aware of the importance of email encryption. Particularly, we found that users are very concerned about identity theft, as 78% want to make sure that no other person is able to write email in their name.

**Keywords:** mail encryption, user study, usability, PGP, S/MIME, pEp.

## 1 INTRODUCTION

To prevent cyber-crime and to protect user privacy, almost all services running on the Internet critically depend on cyber security measures. Most dominantly, transport layer security (TLS) is used for securing a variety of communication protocols. Particularly when browsing the World Wide Web, transport security has found wide adoption and awareness of its imperative necessity. Internet Banking, shopping on Amazon.com, accessing governmental e-services - those are just a few examples where users became more and more aware of the security risks of web applications. A huge step towards more secure Internet communication has been the integration of end-to-end cryptography in mobile internet messenger services such as Whatsapp, Signal or Telegram. In contrast, for securing one of the most commonly used communication channels, the email end-to-end encryption is only applied by a negligible faction of email users [1]. Standardized technologies for cryptographically securing email exchanges have been available for decades. Nevertheless most users rely on unencrypted and unauthenticated email communication, often without being aware that there exist mechanisms, which would mitigate the security implications that come with it. Actually, two major end-to-end encryption technologies exist since decades, namely Pretty Good Privacy (PGP) [2] and Secure Multipurpose Internet Mail Extensions (S/MIME) [3]. A recent initiative called Pretty Easy Privacy (pEp) [4] made efforts to simplify the usage of end-to-end cryptography in email communication for novice users. Unfortunately, those technologies are still barely deployed. According to [1] more than 95% of the overall email traffic is exchanged without end-to-end encryption. Therefore, two main research questions came to our mind: 1) why are users hesitating to use e-mail end-to-end encryption technologies and 2) which usability issues exist that hinder users from securing their daily email communication using end-to-end encryption. To address these questions we have conducted an online survey and user testing in which participants actively use encryption, in order to get a precise and authentic view on usability issues.

The rest of the paper is organized as follows. Section 2 gives an overview about the methodology used to conduct the usability study. Section 3 discusses the obtained results. Finally, Section 4 concludes this paper and gives an overview about future works.

## 2  METHODOLOGY

In this section, we present our approach for evaluating the usability of PGP, S/MIME and pEp.

Before conducting our study, we have identified the most commonly Mail User Agents (MUA) also known as email programs, that - natively or by additional plugins - support at least one of the three technologies PGP, S/MIME or pEp. We assessed the usability of the encryption features in each of these mail programs to get a personal impression as well as to anticipate the challenges that other users might face when cryptographically securing their emails. Actually, we tested the integration of PGP, S/MIME and pEp in today's most commonly used mail programs (MUA) that support end-to-end encryption to: 1) identify which encryption technology is supported by which MUA, 2) prevent participants from testing MUAs that turn out to be unusable (e.g. due to discontinued development, incompatibility of versions and operating system, ...), 3) anticipate the challenges users could face when trying to use these three technologies in the context of a specific MUA to help them to overcome common pitfalls that would otherwise ultimately hinder them from sending a secure e- mail.

Table 1 represents the Mail User Agents considered in our analysis. It also depicts the plugin required to add PGP, S/MIME or pEp functionality to a MUA if not supported natively. From this collection of MUAs, we had to choose the subset of MUAs that will be used for user testing taking into account the popularity of the MUAs and considering that each encryption technology should be tested on each major platform (if supported) and should be free of costs for our participants. We assumed that even if the technology is usable, its implementation within an e-mail program might make it difficult to use and vice-versa. Therefore, we wanted our participants to test two different implementations of e-mail encryption, to have a direct comparison of their usability: a participant would either test two different technologies or she/he would test two different implementations of the same technology. Particularly, when testing two different technologies, we let participants test a pEp implementation and a PGP or S/MIME implementation, in order to see if pEp indeed meets its goal of simplifying the mail encryption process.

**Table 1.** Commonly used mail user agents (MUA) and their support for PGP, S/MIME and pEp

| Technology | Mail User Agents | Plugin | Tested |
|---|---|---|---|
| PGP | Outlook Desktop 2013/2016 | Gpg4o | ✓ |
| | Thunderbird | Enigmail | ✓ |
| | Gmail(Webmail) | FlowCrypt | ✓ |
| | Other Webmail | Mailvelope | ✓ |
| | Apple iOS | iPG mail app. | x |
| | Android | Maildroid and Cryptoplugin | ✓ |
| | Windows Mail | Not Supported | x |
| | Apple Mail (MacOS) | Not Supported | x |
| S/MIME | Outlook Desktop 2013/2016 | Native support | ✓ |
| | Thunderbird | Native support | ✓ |
| | Apple iOS | iPhone mail app | ✓ |
| | Android | Maildroid and Cryptoplugin | ✓ |
| | MacOS | Native support | ✓ |
| | Gmail(Webmail) | Not Supported | x |
| | Other Webmail | Not Supported | x |
| | Windows Mail | Native support | x |
| pEp | Thunderbird | Enigmail | ✓ |
| | Android | Official pEp app | ✓ |
| | Apple iOS | App coming soon | x |
| | Outlook Desktop 2013/2016 | pEp for outlook | x |
| | MacOS | Not supported | x |
| | Gmail(Webmail) | Not Supported | x |
| | Other Webmail | Not Supported | x |
| | Windows Mail | Not supported | x |

To assess the use of PGP, S/MIME and pEp, we have employed two methods: an online survey with 50 participants and conducted a user testing with 12 participants. The details about these studies and the corresponding results are given hereafter.

### 2.1 Online Survey

The aim of the online survey on email end-to-end encryption technologies was three-fold. First, to explore users understanding and awareness of security in emails ex-changes, their expectations and opinions on end-to-end encryption. Second, to learn about the propagation of PGP, S/MIME and pEp. Third, to compare the results of the online survey, which are quantitative, with the results of the user testing, that are qualitative. The survey included closed-ended questions (multiple choice questions), open-ended questions and ranked questions with a balanced rating scale.

### 2.2 User Testing

The user testing was conducted adhering to a predefined testing protocol. Each user testing started with a short interview of the participant, determining some demographic data (age, nationality, profession), the preferred MUA to access her/his emails and previous knowledge the participant had about cryptography in general or email encryption in particular. When the participant was familiar with one of the MUAs, we proposed her/him the test scenario related to this MUA so that she/he could focus on configuring and using encryption features rather than struggling with an unknown mail software. For the participants who did not have any experience or knowledge on any of the MUAs proposed, we helped them to install and set up a MUA up to the point that they were able to successfully access their mail account. Each participant was then asked to enable and configure the security features of the chosen MUA to use a specific email encryption technology and send a secured email to us. When the participants were struggling for more than 10 minutes with a specific configuration step, we helped them. The user test was completed once we received an email sent by the participant that was successfully encrypted and signed.

## 3 EVALUATION

### 3.1 Online Survey Results

The online survey was launched on 30 November 2018 and reached 50 participants on 12 December 2018 when we started the analysis of the results.
The survey began with a demographic section. The majority of the participants was under 30 years old, coming from Germany, Egypt and Morocco. Most of them were students and employees working for IT organizations.
The results concerning the participants personal experience with email exchange showed that emails constitute a remarkable portion of their daily communications, reaching at least 7 emails per day, but most of them were nor encrypted neither signed, 38% received at least 1 mail encrypted per day, and less than half of the participants were obliged to use end-to-end encryption by their organizations. Regarding the use of email software, the participant used more than one software. More than half of the participants used dedicated mobile applications, 50% used webmail, and 44% used dedicated desktop applications.

**Results for PGP**

The results regarding the use of PGP are:
- 60% of the participants never heard about PGP, 40% knew PGP but only 24% were also using it.
- 70% of the participants stated that they could not use PGP for all emails due to the fact that the recipient did not use PGP.
- 25% of the participants thought that it was difficult to find the recipient's public key, 20% thought that configuring PGP was time consuming and just 5% declared that PGP is not implemented on their favorite platform / email client.
- 20% of the participants were always verifying the fingerprint of the recipient key, 30% were doing it occasionally, 35% never did and 15% did not know.

- The participants conceded that PGP guaranties privacy, confidentiality, authenticity and integrity, in addition to the fact that there was no cost for using it. However, they stated that comparing fingerprints was difficult and time consuming, and required the recipient to use it as well, which was not always the case given that PGP was not widely adopted.
- Participants suggested to make PGP supported on all platforms and simplify fingerprint comparison.

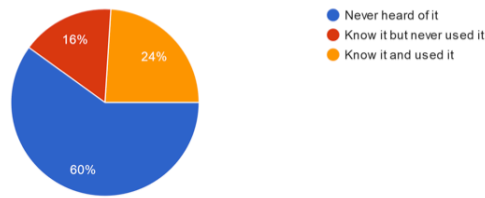

**Fig. 1.** Do you know a technology calle Pretty Good Privacy (PGP) ?

**Results for S/MIME**

The results regarding the use of S/MIME are:
- 64% of the participants never heard about S/MIME, 36% knew it but only 18% were also using it.
- 61% of the participants stated that the recipient was not using S/MIME.
- 28% did not trust digital certificates or its issuing entity and only 11% did not know how to obtain digital certificate.
- 17% encountered difficulties configuring their environment to use S/MIME.
- 27% admitted that they had issues with untrusted certificates and 28% indicated that having to pay for a trustworthy certificate is an obstacle.
- The participants agreed that S/MIME had the advantage of being integrated in most email clients including Apple MacOS/iOS, but they discredited it because they needed to pay to obtain a trustfully certificate.

**Results for pEp**

Regarding pEp, the results showed that it is not as known as PGP and S/MIME and only 10% knew it. No participant stated that she/he ever used it. Moreover, 40% of the participants hesitated to use pEp because their recipients would not use it.

**Results on the overall impression of the users on end-to-end encryption**

The goal of the last part of the survey was to gather the overall impression on end-to-end encryption, by scaling the degree of awareness of the participants on matter of email exchange security, especially if they had an email piracy issue.

Assessing their overall impression, the participants were mostly aware of the importance of email encryption: 66% thought that email encryption is important to very important (34% for important and 32% for very important) .

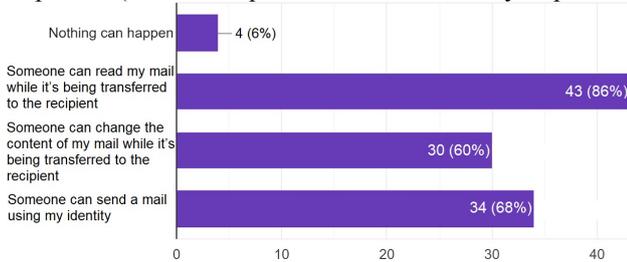

**Fig. 2.** Considering a scenario of using non- encrypted email communication, which of the following may occur ?

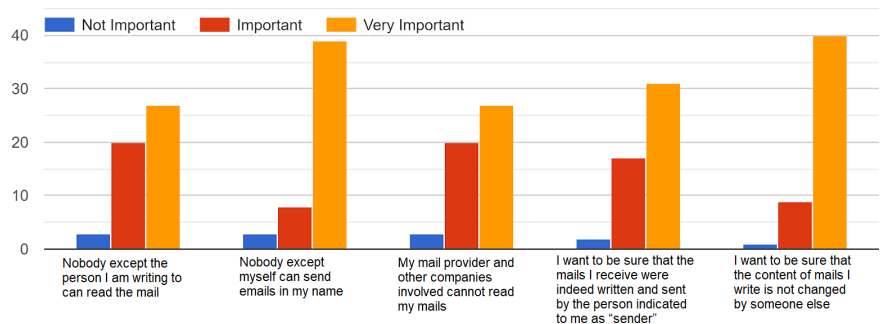

**Fig. 3.** Please indicate the importance that the following security goals have for you

Considering the scenario of non-secured email exchange, more than 60% of the participants could imagine that their emails can be passively or actively tampered with; and even larger percentage of 86% assumed that an entity other than the email recipient can read the email content (Figure 2).

Assessing the importance of specific security goals, almost all of the participants estimated the confidentiality, integrity and authenticity of their emails as important or very important. For only 6% of the participants, confidentiality does not matter and for only 2% the integrity of the sent emails does not matter (Figure 3).

### 3.2 User Testing Results

For the user testing, we have chosen a convenience sample where we have recruited 12 users mostly young people (students and IT users from Germany, Morocco and Egypt but also some family member), between 20 to 30 years old. The user testing has been done twice for each participant to test either two different implementation of the same technology or two different technologies.

The user testing allowed us to identify exactly at which steps users struggle the most. We noticed that the usability of an encryption technology largely depends on its implementation within a mail program, which leads to the fact that the same technology can be laborious to use in one MUA but convenient in another MUA.

The results presented in this part focus on the tasks that we identified as highly difficult for the participants to configure and use a specific encryption technology in the context of a specific mail program.

**Results for PGP**

The participants faced several difficulties depending on the MUA used:

- **Outlook 2016**: For the task "upload key pair", the users were not able to import an existing key-pair. The task "Get recipient key" was also difficult because the default configuration of Gpg4o is to connect to a key-server using an unusual port, which sometimes results in an empty response or a refused connection. The users were not able to find and download the recipient public key on this key server. Actually, is not trivial for users to identify this issue and accordingly go into the respective Outlook settings to change the port number to the one that is commonly used in combination with that key server. Moreover, when composing a new secure email and opting-in encryption and/or signature feature(s), the graphical user interface became distorted. Buttons, labels and text fields were misaligned and overlapped, making the user interface almost unusable particularly difficult to assess whether the correct option was selected.
- **Thunderbird**: Thunderbird required not only installing Enigmail plugin, but also extra configuration options that are not easy to find for the users. The participants had to activate the option "Force using PGP" in the privacy settings of Thunderbird after installing the plugin. Moreover, the option "Activate PGP" had to be applied for each mail account in the account settings. Finally, the step "Get recipient key" was identified as difficult, because Enigmail searched for the missing recipient public keys on only one server at a time. It was up to the users to manually change the key server on which Enigmail searches for missing keys, which required patience and willingness to tediously change the key server settings until succeeding in retrieving the recipient public key from one of the servers on which the recipient published her/his key.
- **Maildroid**: Maildroid (as all other mobile apps analyzed) offered no functionality to generate a PGP key-pair directly on the mobile device. The key had to be generated externally (e.g. on a PC) and then be transferred back to the mobile device. There were multiple ways to transfer the generated key pair (e.g. self-sending it via email, upload to a cloud, USB exchange), but it required intense user interaction and downgraded the user experience.

As summary, we recognize that PGP requires many configuration steps until successful usage, which was particularly the case for the task concerning the import of the public keys of recipients. This task always turned out to be difficult or tedious for all the participants, regardless of the tested platform. This is due to the design principle of PGP to let full control to the users with respect to key management – which at the same time is demanding a basic understanding of asymmetric cryptography and the technology. Following the user testing for PGP, we could conclude that the most difficult MUA to use PGP with was Thunderbird, because of the difficulties that the users faced in addition to the fact that the buttons to get configuration steps done are hidden deeply in setting menus. In contrast, FlowCrypt was the easiest tool to use PGP with, as it generates a key-pair for new user with only a few clicks and searches for the recipient key automatically on almost all commonly used key-servers. Thereby FlowCrypt solves nearly all usability issues encountered by the participants. Unfortunately it comes with two downsides: FlowCrypt uploads the generated key-pair only on its proprietary key server which is unknown to most other PGP implementations, thus making the import of public keys of FlowCrypt users harder for other users. Secondly, up to now, FlowCrypt only supports Gmail webmail.

**Results for S/MIME**
The difficulties encountered by the users are as follows:

- **Outlook 2016/2013**: The configuration option to let the users import their own digital certificates was not easy to find and the participants passed too much time looking for the button in the settings to import their certificate. In addition, they experienced a strange bug. The users could encrypt their outgoing emails only when replying to an encrypted email already received, but they could not encrypt a new email even though they already retrieved the certificate of the recipient.
- **iOS 12**: The users had to decompress, on a computer, the archive containing the requested certificate, received by email, from the issuing Certification Authority. Then, they had to send back the certificate (the .pfx file) as an attachment in an email to themselves. In addition, before importing the certificate, they needed to activate S/MIME manually in the settings of the iPhone. However, the respective setting option was fairly hidden in the phone settings menu.
- **Android**: The users had to decompress the received certificate file using another platform (e.g. a PC). Then, transfer back the certificate to the mobile device after decompressing it. It was the same problem as for iOS. They could only transfer it by sending an email to themselves containing the certificate (the .pfx file).

To conclude, thanks to the way S/MIME works, it can be used easily by novice users, because the users do not have to generate any keys. They receive both public and private digital certificate and they have just to import it into the MUA. Once a user receives a signed email, the sender's public certificate is integrated automatically into the MUA. So, the users do not have to do any supplementary tasks other than configuring S/MIME in the desired MUA. However, we can conclude that is very difficult to use S/MIME on Outlook, as the options to import the digital certificate are difficult to find and the user can send an encrypted email only as reply to an email that is already secured via S/MIME (digitally signed). Also, on iOS it is very difficult to activate S/MIME on the device as the option is not easy to find and there is not any way for the users to send the pfx file back to the device than to send an email to themselves containing the pfx file. Moreover, the configuration to activate S/MIME on the iOS devices varies from one iOS version to another.

**Results for pEp**
pEp required only few tasks in order to configure it and use it compared to PGP and S/MIME. Thanks to automated key management, non-technical wording in its user interfaces and abstraction of security features, pEp did not show any major usability issues that would hinder (novice) users from using it. The comparison of trustwords through the so-called pEp handshake, in order to establish trust in the recipient key, was considered as convenient and rather an easy task to do by most of the participants. Nevertheless, some of the participants did not understand why the handshake was necessary and what to do with the trustwords shown during the handshake graphical user interface. pEp showed to be the easiest technology to use, but unfortunately being not (yet) compatible with all major platforms. As a consequence, we could not test it on Apple MacOS or Apple iOS platforms, which was used by a large fraction of email users.

## 4      RELATED WORK

In this section, we will give a brief overview on related work. In 2012, Moecke and Volkamer analyzed all different email services, defining security, usability and interoperability criteria and applied them to existing approaches. Based on the results, closed and web-based systems like Hushmail were more usable, contrarily to PGP and SMIME that require add-ons to carry the key in a secure way [17]. In 2017, Lerner, Zeng and Roesner from University of Washington, presented a case study with people who frequently conduct sensitive business. They estimated the confidence put on encrypted emails using a prototype they developed based on Keybase for automatic key management [18]. In 2018, Clark et al. conducted a study focused on: 1) systematization of secure email approaches taken by industry, academia, and independent developers; 2) evaluation for proposed or deployed email security enhancements and measurement of their security, deployment, and usability. Through their study, they concluded that deployment and adoption of end-to-end encrypted email continues to face many challenges: usability on a day-to-day scale; key management which remains very unpractical.[19]. In 2018, a group of researchers from Brigham Young University and University of Tennessee conducted a comparative usability study on key management in secure emails tools, in which they oversaw a comparative based study between passwords, public key directory (PKD), and identity-based encryption (IBE). The result of the study demonstrated that each key management has its potential to be successfully used in secure email [20].

## 5   CONCLUSION

In this paper, we identified the most frequent usability issues that users face when protecting their email communication using PGP, S/MIME or pEp. Using both online survey and user testing, we had an overall view on the awareness of the users on email encryption as well as a detailed view on the difficulties they can encounter. These difficulties have certainly an impact on the fact that they hesitate to use PGP, S/MIME or pEp.
Thanks to the online survey, we were able to identify the usability issues of each technology and assess the general impression of our audience towards the importance of email encryption. The results of the online survey showed us that the users were aware of the importance of email encryption with 32% saying it is very important. Additionally, users were very concerned about identity theft, as 78% of the participants wanted to make sure that no other person is able to write an email using their name and 80% of the participants wanted to be sure that the content of their mail is not changed by someone else, while being transferred to the recipient. This result shows that for many users, signing emails is more important than encrypting them.

Currently, we are finalizing the correlation of the online survey answers with the results of the user testing, to validate if the participants of the online survey have the same usability issues as the participants of the user testing while using a certain technology. For future work, we plan to conduct more user testing with different kinds of people (people with different age categories). In addition, thanks to the results obtained regarding identity theft and to some feedback we had on our study, we will set up another online survey to know the measures (if any) taken by the users to protect their identity.